\documentclass[sigconf]{acmart}
\AtBeginDocument{%
  }

\acmConference[AGENT 2026]{International Workshop on Agentic Engineering}{Apr.~14,
  2026}{Rio de Janeiro, Brazil}

\usepackage{subcaption}
\usepackage{enumitem}
\usepackage{multirow}
\usepackage{algorithm}
\usepackage{algpseudocode}
\usepackage[table]{xcolor}
\usepackage{listings}
\usepackage{xcolor}

\newcommand{\remark}[1]{}
\newenvironment{hide}[1]{\remark{#1}}{}

\newcommand{\full}[1]{}                 

\begin{document}

\title[TALM: Dynamic Tree-Structured Multi-Agent Framework]{TALM: Dynamic Tree-Structured Multi-Agent Framework with Long-Term Memory for Scalable Code Generation}



\author{Ming-Tung Shen}
\email{r12725051@ntu.edu.tw}
\author{Yuh-Jzer Joung}
\orcid{0000-0003-0015-1258}
\email{joung@ntu.edu.tw}
\affiliation{
 \institution{Dept.~of Information Management, National Taiwan University }
 \city{Taipei}
 \country{Taiwan}
}

\begin{hide}{
\author{Lars Th{\o}rv{\"a}ld}
\affiliation{%
  \institution{The Th{\o}rv{\"a}ld Group}
  \city{Hekla}
  \country{Iceland}}
\email{larst@affiliation.org}

\renewcommand{\shortauthors}{Trovato et al.}
}\end{hide}

\begin{abstract}
%
Agentic code generation requires large language models (LLMs) capable of complex context management and multi-step reasoning. Prior multi-agent frameworks attempt to address these challenges through collaboration, yet they often suffer from rigid workflows and high reasoning recovery costs. To overcome these limitations, we propose 
TALM (Tree-Structured Multi-Agent Framework with Long-Term Memory),
a dynamic framework that integrates structured task decomposition, localized re-reasoning, and long-term memory mechanisms. 
TALM employs an extensible tree-based collaboration structure. The parent–child relationships, when combined with a divide-and-conquer strategy, enhance reasoning flexibility and enable efficient error correction across diverse task scopes.
Furthermore, a long-term memory module enables semantic querying and integration of prior knowledge, supporting implicit self-improvement through experience reuse. Experimental results on HumanEval, BigCodeBench, and ClassEval benchmarks demonstrate that TALM consistently delivers strong reasoning performance and high token efficiency, highlighting its robustness and practical utility in complex code generation tasks.
\end{abstract}

\ccsdesc[500]{Computing methodologies~Multi-agent systems}

\keywords{Large Language Model, AI Agent, Multi-Agent Framework, Code Generation, Tree-Structured Collaboration}


\maketitle

\section{Introduction}
Large Language Models (LLMs) have demonstrated strong capabilities in natural language processing and, in particular, code generation, where both pre-trained and fine-tuned models can efficiently produce accurate code \cite{wang-etal-2021-codet5, li2023starcoder, luo2024wizardcoder}. Recent efforts have sought to further improve performance through prompting frameworks, such as chain-of-thought \cite{wei2022chain} or self-planning reasoning \cite{jiang2024self}, which guide LLMs to follow predefined reasoning path. While effective for short functions and modular tasks, these approaches rely on a single reasoning path and struggle when requirements become complex or context exceeds the model’s window size \cite{levy-etal-2024-task, li-etal-2024-loogle}. 

To address longer and more ambiguous tasks, multi-agent frameworks have emerged, leveraging role-playing and workflow decomposition to mimic software development life cycle \cite{hong2024metagpt, qian2024chatdev}. However, most prior multi-agent systems still adopt rigid pipelines, which makes them difficult to adapt to varying software requirements.

Moreover, due to the hallucination problem \cite{Ji2023hallucination} inherent in LLMs, the reasoning process cannot always be considered reliable. An effective re-reasoning mechanism is thus essential for ensuring robust performance. However, existing multi-agent frameworks either do not explicitly check for reasoning flaws, allowing errors to propagate downstream \cite{islam-etal-2024-mapcoder}, or they resort to restarting the entire process when mistakes are detected, leading to unnecessary resource consumption \cite{wang2025tdag}. In our view, localized error correction is more appropriate for code generation tasks, since code defects typically stem from partial solutions rather than a complete misdirection.

To overcome these limitations, we propose \textbf{TALM} (\textbf{T}ree-structured multi-\textbf{A}gent framework with \textbf{L}ong-term \textbf{M}emory). TALM organizes agents in a tree structure to enable flexible divide-and-conquer task decomposition and supports localized re-reasoning for efficient error correction. When the system receives a user’s task prompt, the task is decomposed into multiple subtasks, and the dependencies among these subtasks are represented as a tree structure. This design allows TALM to handle tasks of different granularity. Furthermore, the tree structure enables TALM to perform re-reasoning within a specific subtree when flaws are detected. 

In addition, to achieve lifelong learning similar to humans, TALM integrates a long-term memory module that stores prior experiences and retrieves relevant knowledge during solving new tasks, enabling implicit self-improvement. By leveraging past successes and avoiding previously observed mistakes, TALM gradually evolves over time without additional retraining.

We evaluate TALM on HumanEval \cite{chen2021evaluating}, BigCodeBench \cite{zhuo2024bigcodebench}, and ClassEval \cite{du2023classeval} benchmarks. Results show that TALM outperforms existing approaches while maintaining high token efficiency. Beyond benchmark results, TALM provides a general paradigm for integrating structured reasoning with long-term memory, suggesting a path forward for scalable, reliable and self-improving code generation agents.


\section{Related Work}
\label{sec:related}
Recent advances in prompting have enhanced the reasoning ability of LLMs without requiring model retraining. Basic strategies include zero-shot prompting, where models are asked to directly produce outputs from task descriptions, and few-shot prompting, where a small set of exemplars helps align input-output mappings \cite{brown2020language}. Building on this, chain-of-thought prompting \cite{wei2022chain} and self-planning prompting \cite{jiang2024self} encourage models to externalize intermediate reasoning steps, improving performance on multi-step problems. ReAct \cite{yao2023react} further integrates reasoning and acting by allowing LLMs to decide when to deliberate internally and when to invoke external tools, thereby improving tool usage in complex tasks. More recently, Reflexion \cite{shinn2023reflexion} introduces a self-reflective loop that enables models to critique and revise their own outputs, while tree-of-thought \cite{yao2023tree} organizes reasoning as a search process over a tree structure, enabling exploration of multiple reasoning paths. These approaches have proven cost-efficient compared to fine-tuning, and they extend LLM capabilities to increasingly complex reasoning tasks.


Multi-agent frameworks have recently emerged as a promising direction for enhancing LLM-based reasoning and task execution. By decomposing tasks into subtasks and assigning them to role-specific agents, these frameworks emulate the division of labor in human teams and provide a way to overcome the context-window limitations of single-agent approaches.

AutoGen \cite{wu2023autogen} provides a general multi-agent platform where a centralized planner coordinates multiple specialized agents through role-playing and structured dialogues. This design improves modularity and allows flexible integration of external tools.
TDAG~\cite{wang2025tdag} dynamically decomposes tasks into executable steps and instantiates specialized sub-agents as needed. Its reflection mechanism enables feedback and backtracking, enhancing robustness and recovery.
AgentNet~\cite{yang2025agentnet} employs a dynamically evolving DAG communication topology, allowing agents to adapt roles and strategies through an evolution phase that fosters specialization and autonomous task allocation.
In contrast, ChatDev \cite{qian2024chatdev} focuses on simulating the software development life cycle by assigning agents to roles such as analyst, designer, coder, and tester, with dedicated mechanisms for cross-stage context menagement. MetaGPT \cite{hong2024metagpt} emphasizes responsibility separation: each agent operates within well-defined input–output boundaries, and communication is achieved through a subscription-based messaging mechanism. 

Beyond workflow simulation, MapCoder \cite{islam-etal-2024-mapcoder} introduces a retrieval agent that augments code generation with relevant exemples. By dynamically retrieving semantically similar problem–solution pairs from past knowledge, MapCoder enhances few-shot prompting and improves performance. 



Despite advances in multi-agent frameworks for code generation, challenges remain—namely rigid workflows, high reasoning correction costs, and the absence of cumulative memory. TALM is proposed to address these limitations.




\section{TALM Framework}
\label{sec:framework}

TALM is designed to enhance code generation through a dynamically growing tree structure, localized error correction, and knowledge reuse. The framework consists of three main components: (1) a tree-structured collaboration mechanism, (2) a localized re-reasoning process, and (3) a long-term memory module. Together, these components enable agents to coordinate flexibly, correct reasoning errors efficiently, and accumulate experience for future tasks.

\begin{figure*}[t]
  \centering
  \includegraphics[width=0.75\textwidth]{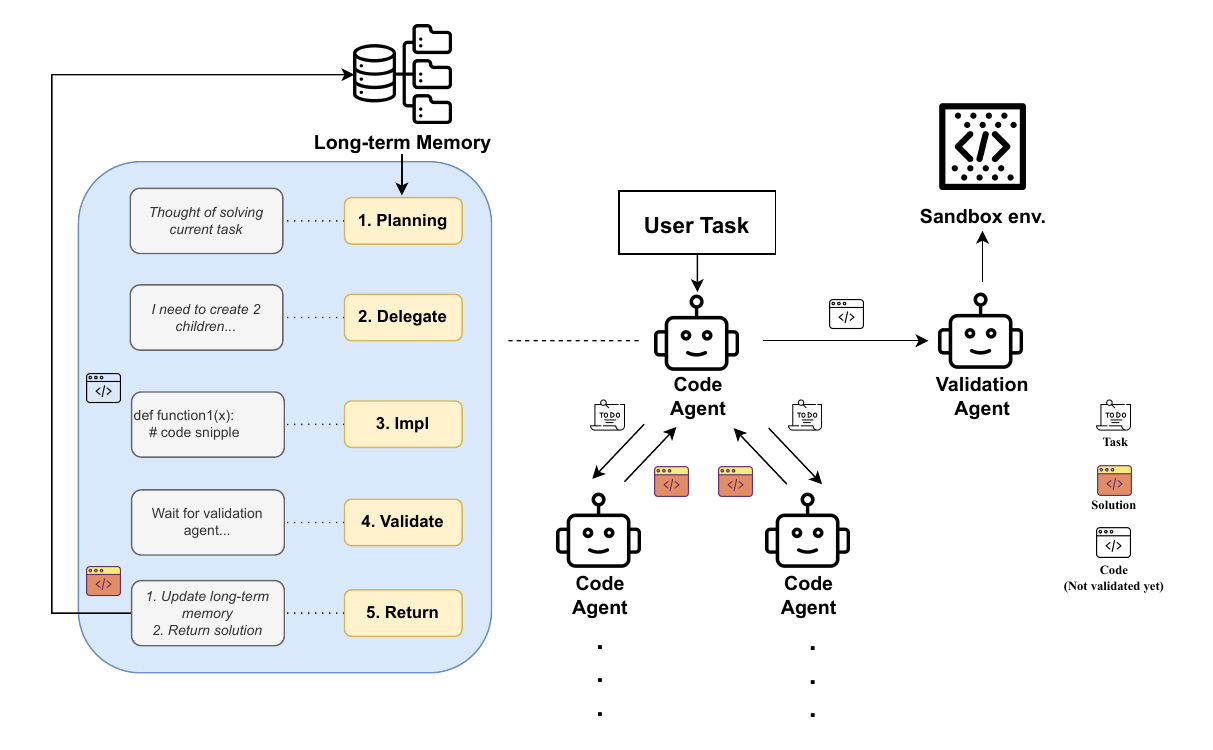}
  \caption{TALM workflow overview. The root agent decomposes the task into a tree of subtasks; child agents solve localized steps and return results upward.}
  \Description{A page-wide diagram showing a root agent splitting a task into a hierarchical tree; local verification triggers subtree-only reruns; logs flow into a memory store.}
  \label{fig:overview}
\end{figure*}

Figure~\ref{fig:overview} illustrates the overall workflow. A user’s task is recursively decomposed into subtasks handled by specialized code agents. Each agent attempts to solve its assigned subtask based on the given context and then returns the result to its parent. Before returning, the output is checked by the validation agent, which generates test cases and executes them in a sandbox environment. During processing, agents can also query the long-term memory module to retrieve past experiences as external knowledge. In this way, TALM decomposes the original task into finer-grained subtasks, ensures that each agent has a clear and focused context, and produces the final output through a bottom-up merging process along the tree.

\begin{small}{
\begin{algorithm}[h]
\caption{TALM Workflow}
\label{alg:talm-workflow}
\begin{algorithmic}[1]
\Statex \texttt{// Initialize global control parameters}
\Statex \texttt{// These values can be configured as hyperparameters}
\Statex \texttt{// m: maximum tree height}
\Statex \texttt{// n: initial tree branching degree}
\Statex \texttt{// k: decay rate for degree as depth increases}
\Statex \texttt{// r: maximum number of verification retries}
\State $m \gets 3$
\State $n \gets 3$
\State $k \gets 1$
\State $r \gets 3$
\vspace{0.75em}
\Procedure{ExecuteWorkflow}{$user\_instruction$}
    \Statex \hspace{\algorithmicindent} \texttt{// Create root code agent}
    \State $finalCode \gets$ \Call{ExecuteCodeAgent}{$user\_instruction$, $1$, $n$}
    \State \textbf{return} $finalCode$
\EndProcedure
\vspace{0.75em}
\Procedure{ExecuteCodeAgent}{$task$, $height$, $degree$}
    \State $plan \gets$ \Call{Analyze}{$task$} 
    
    \If{$height < m$}
        \Statex \hspace{\algorithmicindent} \hspace{\algorithmicindent} \texttt{// Make sure $|Subtasks| \le degree$}
        \State $subtasks \gets$ \Call{Decompose}{$plan$, $degree$}
    \Else
        \State $subtasks \gets \emptyset$
    \EndIf

    \State $child\_results \gets [\ ]$
    
    \ForAll{$subtask$ \textbf{in} $subtasks$}
        \Statex \hspace{\algorithmicindent} \hspace{\algorithmicindent} \texttt{// Create child code agent}
        \State $result \gets$ \Call{ExecuteCodeAgent}{$subtask$, $height + 1$, $degree - k$}
        \State \Call{Append}{$child\_results$, $result$}
    \EndFor

    \State $code \gets$ \Call{Implement}{$plan$, $child\_results$}
    \State $verifiedCode \gets$ \Call{ExecuteVerificationAgent}{$code$}
    \State \textbf{return} $verifiedCode$
\EndProcedure
\vspace{0.75em}
\Procedure{ExecuteVerificationAgent}{$code$}
    \State $retry \gets 0$
    \Repeat
        \State $testResult \gets$ \Call{RunTests}{$code$}
        \If{$testResult \ne \text{Pass}$}
            \State $code \gets$ \Call{FixErrors}{$code$, $testResult$}
            \State $retry \gets retry + 1$
        \EndIf
    \Until{$testResult = \text{Pass}$ \textbf{or} $retry = r$}
    \State \textbf{return} $code$
\EndProcedure
\end{algorithmic}
\end{algorithm}
}\end{small}

\subsection{Tree-Structured Collaboration}

In TALM, agents are organized as nodes in a tree that mirrors the decomposition of the original task. Higher-level nodes oversee broader objectives, while lower-level nodes are responsible for finer-grained subtasks. Each agent can either (i) attempt to complete its subtask independently or (ii) delegate part of the work by spawning child agents. The results returned by children are always passed through the validation agent, which generates test cases and executes them in a sandbox environment to ensure reliability, before being integrated by the parent. Verified outputs may serve as helper functions or partial solutions that support the parent’s own task. The final solution emerges through a bottom-up merging process, closely resembling a divide-and-conquer strategy.

More specifically, each code agent in the tree proceeds through the following process:
\begin{itemize}[leftmargin=*]
  \item \textbf{Planning Phase:} The code agent formulates a plan for solving the assigned task, drawing on the task description and relevant knowledge retrieved from long-term memory. The output of this phase is a step-by-step plan to guide subsequent stages. 
  \item \textbf{Delegation Phase:} If the plan indicates that the solution requires multiple modules or distinct functions, the code agent designs subtask descriptions, spawns child agents, and integrates their returned solutions into its context. If delegation is unnecessary, this phase is skipped.
  \item \textbf{Implementation Phase:} The code agent implements the solution code using the accumulated context.
  \item \textbf{Validation Phase:} The generated code is passed to the validation agent, which ensures executability and correctness by running test cases in a sandbox environment. If errors are detected, the validation agent attempts debugging for up to $r$ iterations.
  \item \textbf{Return Phase:} The code agent updates the long-term memory with its current experience and returns the verified solution code to its parent.
\end{itemize}
Details regarding the interaction with long-term memory are discussed in Section~\ref{sec:memory}, and the overall workflow is summarized in Algorithm~\ref{alg:talm-workflow}.

\begin{figure*}[t]
  \centering
  \begin{subfigure}{0.49\textwidth}
    \centering
    \includegraphics[width=\linewidth]{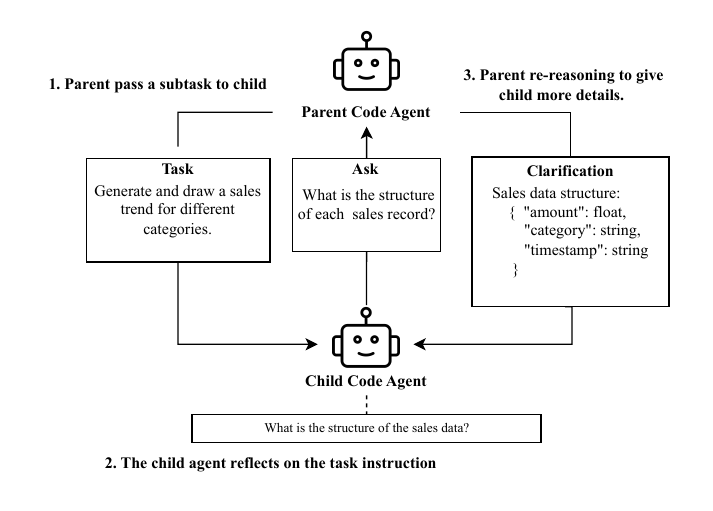}
    \caption{Child-Agent Clarification (bottom-up).}
    \Description{A child agent requests clarifications from its parent when the task specification is ambiguous; the parent refines requirements, enabling correct implementation.}
    \label{fig:clarification}
  \end{subfigure}
  \hfill
  \begin{subfigure}{0.49\textwidth}
    \centering
    \includegraphics[width=\linewidth]{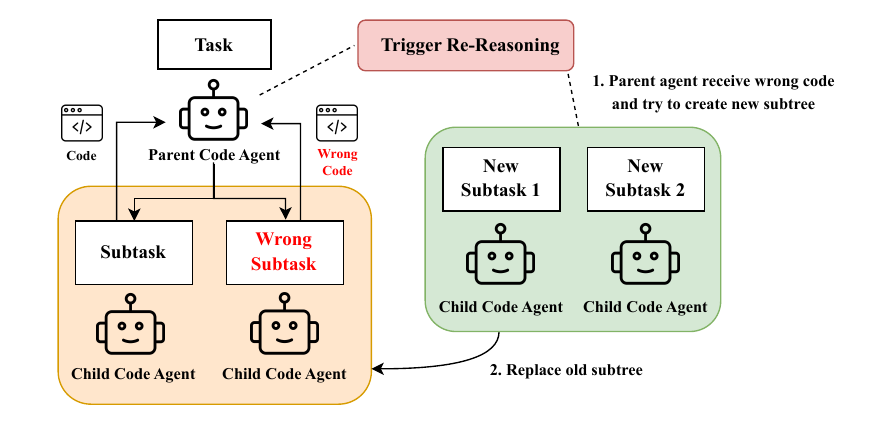}
    \caption{Structure Correction (top-down).}
    \Description{The parent agent identifies a flawed decomposition and issues a corrected structure; only the affected subtree is recomputed.}
    \label{fig:structure}
  \end{subfigure}
  \caption{Localized re-reasoning in TALM. Left: child-to-parent clarification when task specs are underspecified; Right: parent-initiated structure correction when decomposition is flawed.}
  \Description{Two diagrams: (a) shows bottom-up clarifications; (b) shows top-down correction and targeted subtree recomputation.}
  \label{fig:localized}
\end{figure*}

To prevent uncontrolled tree growth, TALM introduces three hyperparameters: the maximum depth $m$, the initial branching factor $n$, and the decay rate $k$ ($m, n, k \in \mathbb{Z}^+$). A node may create children only if its depth is less than $m$, ensuring that nodes at depth $m$ are leaves. The number of children a node may spawn begins at $n$ and decreases by $k$ at each subsequent level, producing a narrowing tree shape. This decay mechanism encourages deeper nodes to concentrate on highly specific subtasks, while also constraining overall complexity. Together, these parameters regulate the height and width of the collaboration tree, striking a balance between flexibility, resource efficiency, and structural controllability.\footnote{The decay mechanism is designed under the assumption that task decomposition progressively reduces problem complexity. In the rare cases where a subtask does not become simpler or may even increase in difficulty during decomposition, the framework still allows the subtask to be handled by a single agent without further branching, relying on the agent’s standalone problem-solving capability.}

\subsection{Localized Re-Reasoning}

Traditional frameworks often re-execute the entire workflow when reasoning errors are detected, leading to substantial correction costs and unnecessary recomputation. TALM instead supports \textit{localized re-reasoning}: only the agents within the affected subtree are recomputed, while the rest of the tree remains intact. This strategy reduces redundant computation, preserves verified reasoning chains, and enhances overall efficiency. By leveraging the hierarchical structure, TALM introduces localized error recovery mechanisms that improve robustness without sacrificing scalability.

Localized re-reasoning in TALM is realized through two complementary modes: clarification-based and structure-correction (Figure~\ref{fig:localized}). These two modes address errors that arise at different stages of the task execution process and together provide a systematic error-handling capability.

\paragraph{Child-Agent Clarification (bottom-up).}
Clarification-based re-reasoning is triggered when a child agent detects that the instructions it received from its parent are underspecified, ambiguous, or incomplete. In such cases, instead of proceeding with insufficient context, the child agent actively requests clarification from its parent. The parent then reflects on its own reasoning trace and provides refined specifications. This feedback loop occurs early in the reasoning pipeline, preventing small ambiguities from propagating downstream. 

\paragraph{Structure-Correction (top-down).}
Structure-correction re-reasoning occurs later in the process, after a parent agent has collected and reviewed the outputs of its children. If the parent determines that its initial task decomposition was flawed—such as assigning irrelevant subtasks or overlooking critical dependencies—it revises its plan and replaces the entire subtree rooted at self with a new set of children and subtasks. The previous results are discarded, and the regenerated subtree reflects a more task-aligned decomposition. This mechanism allows TALM to recover from structural misjudgments in the decomposition process while minimizing wasted effort and avoiding a full workflow restart.

\subsection{Long-Term Memory}
\label{sec:memory}
Most existing multi-agent frameworks are stateless, meaning that agents operate on a one-off basis without retaining knowledge from past experiences. As a result, similar tasks must be repeatedly solved from scratch.
To address this limitation, TALM incorporates a long-term memory module that allows agents to store, retrieve, and reuse knowledge across tasks,
enabling cumulative learning and improved efficiency over time.

\begin{figure*}[t]
  \centering
  \includegraphics[width=0.75\textwidth]{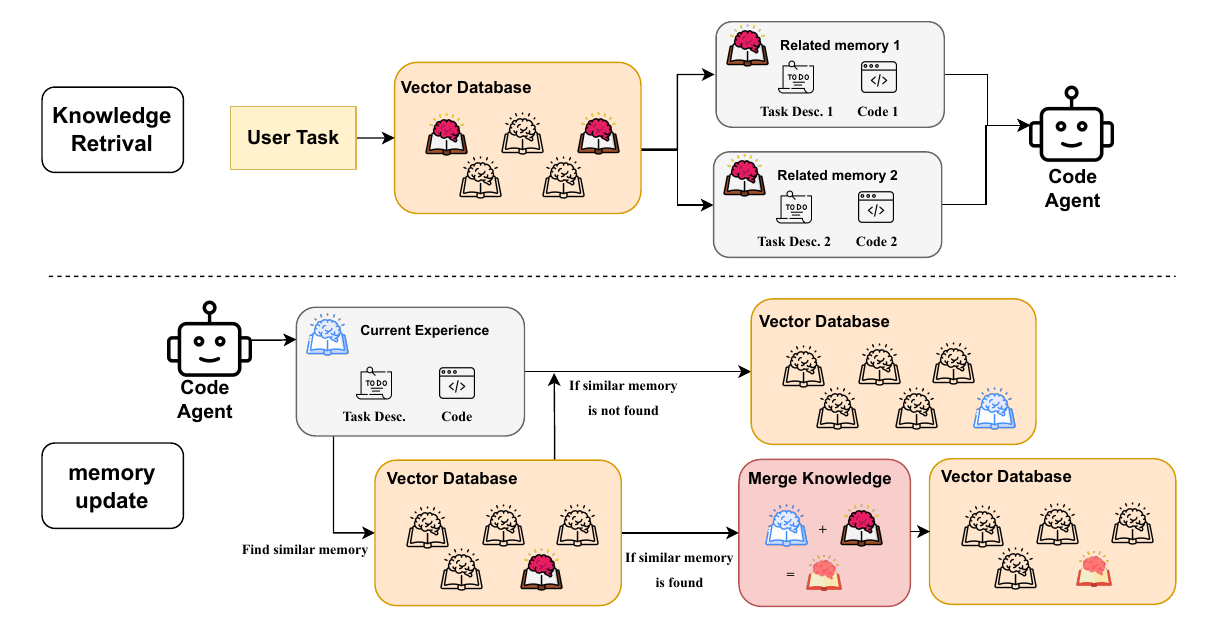}
  \caption{Long-term memory mechanisms in TALM. The top part illustrates knowledge retrieval, where a new task is encoded and matched with similar past experiences in the vector database. The bottom part illustrates memory update, where validated outputs are stored and merged with existing records when overlap is detected.}
  \Description{A diagram with two sections: the top shows knowledge retrieval, where a task description is encoded and searched against past records in a vector database; the bottom shows memory update, where newly validated results are added or merged with existing entries to avoid redundancy.}
  \label{fig:memory}
\end{figure*}

The memory module is built on a vector-based storage system that records task histories, including problem descriptions, reasoning traces, generated code, and structural metadata such as the agent’s tree depth. During new task execution for each code agent, the \textit{knowledge retrieval} mechanism encodes the incoming task description into a semantic vector representation and performs approximate nearest-neighbor search against the stored task vectors in the database. Candidate results are then filtered based on metadata such as tree depth to ensure comparable granularity with the querying agent. This retrieval process provides contextual references that can be incorporated into the current reasoning context. This process is illustrated in the top part of Figure~\ref{fig:memory}.

After the code agent completes its subtask and the output passes validation, the \textit{memory update} mechanism is triggered. The verified code, together with the associated problem description, reasoning process, and metadata, is encoded and stored back into the vector database. Unlike standard retrieval-augmented generation (RAG) approaches~\cite{NEURIPS2020_6b493230,gao2024retrievalaugmentedgenerationlargelanguage} that treat the memory as a static repository for query-time retrieval only, TALM actively maintains and refines its long-term memory through consolidation.

Specifically, the newly encoded entry is compared against existing memory records using vector similarity search. Past entries whose similarity exceeds a predefined threshold (0.95 in our implementation) are retrieved and jointly considered with the new entry. These highly similar records are then provided to LLM, which synthesizes them into a single representative memory entry that unifies their shared knowledge. This synthesis produces updated versions of the problem description, reasoning process, and generated code, capturing the common structure and solution patterns across experiences. The consolidated entry is stored back into the memory, while the original overlapping records are removed. This process prevents uncontrolled memory growth, reduces redundancy, and mitigates context contamination, thereby keeping the memory compact, coherent, and efficient for sustained multi-task reasoning. This update process is illustrated in the bottom part of Figure~\ref{fig:memory}.

By integrating tree-structured collaboration, localized re-reasoning, and long-term memory, TALM balances flexibility, efficiency, and continual improvement, positioning it as a robust framework for complex code generation.  A representative code generation example produced by TALM is provided in Appendix~\ref{app:code-sample}.

\section{Experimental Results}
\label{sec:experimental}
\subsection{Experimental Settings}
\subsubsection{Datasets}
To evaluate TALM across tasks of varying scope and complexity, we adopt three representative datasets: HumanEval \cite{chen2021evaluating}, BigCodeBench \cite{zhuo2024bigcodebench}, and ClassEval \cite{du2023classeval}. HumanEval contains 164 programming problems focused primarily on single-function code generation and serves as a classic benchmark for code synthesis. BigCodeBench includes 1,140 function-level problems with substantially higher difficulty, reflecting more realistic engineering challenges such as data analysis, operating system interaction, and protocol parsing that often require external library usage. ClassEval consists of 100 class-level tasks designed to test object-oriented programming. It emphasizes structural consistency and logical coherence across multiple methods, presenting greater demands on context handling and long-horizon reasoning.

\subsubsection{Baselines}
We compare TALM against a range of strong baselines and representative approaches spanning prompting methods, reasoning strategies, and multi-agent frameworks.  
\textbf{Direct Prompting} serves as the simplest baseline, where task descriptions are directly fed into the model without additional structure, equivalent to a zero-shot setting.  
\textbf{Chain-of-Thought Prompting (CoT)} \cite{wei2022chain} enhances reasoning by encouraging step-by-step explanations before code generation.  
\textbf{Self-Planning Prompting} \cite{jiang2024self} separates problem solving into planning and implementation phases, allowing the model to first determine an overall strategy before synthesizing code.  
\textbf{Reflexion} \cite{shinn2023reflexion} introduces an iterative process where initial outputs are critiqued and corrected based on test feedback, simulating human-like debugging.  
Finally, \textbf{MapCoder} \cite{islam-etal-2024-mapcoder} represents a multi-agent collaboration approach. It augments code generation with a retrieval agent that provides few-shot exemplars, which are then processed by role-specific agents in multiple stages.  

Together, these baselines cover different paradigms—from direct prompting to advanced reasoning and multi-agent collaboration—providing a comprehensive comparison to highlight TALM’s contributions.\footnote{We note that some well-known multi-agent frameworks, such as MetaGPT and AgentNet, are not included as baselines. MetaGPT is excluded because its outputs are difficult to evaluate automatically in our experimental setup, while AgentNet was released after the experiments in this paper were conducted.}

\subsubsection{Foundation Models, Evaluation Metric, \& Hyperparameters}
For the main results reported in Section~\ref{sec:results}, we conduct experiments with both GPT-4o-mini and GPT-4o to ensure a comprehensive comparison across methods. For the ablation studies in Section~\ref{sec:ablation}, we use GPT-4o-mini only, due to cost considerations. In all experiments, the temperature is fixed at $0$ to maintain determinism. Also, for all baseline approaches, we re-implement and evaluate them using the same foundation models (GPT-4o-mini and GPT-4o), rather than directly adopting results from prior work, to ensure a fair comparison.

Due to budget constraints, we limit our experiments to these two models. Under a fixed budget, we argue that comparing methods across different capability tiers of the same model family provides a clearer assessment of generalization, compared to a horizontal comparison across models from different providers (e.g., OpenAI vs. Anthropic) or across open-source and closed-source models, where architectural and training differences may introduce additional confounding factors.

\begin{table*}[t]
\begin{tiny}
\centering
\setlength{\tabcolsep}{8pt}
\renewcommand{\arraystretch}{1.15}
\resizebox{\textwidth}{!}{%
\begin{tabular}{clcccc}
\toprule
& \textbf{Approach \textbackslash\ Benchmark} 
& \textbf{HumanEval} 
& \textbf{BigCodeBench} 
& \textbf{\begin{tabular}[c]{@{}c@{}}ClassEval \\ (testcase level)\end{tabular}} 
& \textbf{\begin{tabular}[c]{@{}c@{}}ClassEval \\ (class level)\end{tabular}} \\
\midrule
\multirow{7}{*}{\textbf{GPT-4o-mini}}
& Direct                 & 87.20\% & 45.20\% & 74.40\% & 33.00\% \\
& CoT                    & 88.45\% & 45.80\% & 76.16\% & 32.00\% \\
& Self-Planning          & 87.07\% & 46.27\% & 75.38\% & 33.00\% \\
& Reflexion              & 86.92\% & 44.43\% & 78.21\% & 31.00\% \\
& MapCoder               & 85.37\% & 41.40\% & 77.22\% & 34.50\% \\
& TALM (without memory)  & \cellcolor[HTML]{FFF2CC}90.59\% & 46.32\% & 77.85\% & 36.00\% \\
& TALM (with memory)     & 89.13\% & \cellcolor[HTML]{FFF2CC}47.10\% & \cellcolor[HTML]{FFF2CC}80.72\% & \cellcolor[HTML]{FFF2CC}36.50\% \\
\midrule
\multirow{7}{*}{\textbf{GPT-4o}}
& Direct                 & 89.10\% & 50.10\% & 80.13\% & 33.00\% \\
& CoT                    & 90.93\% & 49.33\% & 81.91\% & 33.00\% \\
& Self-Planning          & 91.34\% & 51.47\% & 82.66\% & 35.00\% \\
& Reflexion              & 91.70\% & 52.06\% & 84.78\% & 36.00\% \\
& MapCoder               & 93.92\% & 52.32\% & 84.19\% & 36.00\% \\
& TALM (without memory)  & \cellcolor[HTML]{FFF2CC}94.48\% & 52.28\% & 84.65\% & 36.00\% \\
& TALM (with memory)     & 92.55\% & \cellcolor[HTML]{FFF2CC}53.27\% & \cellcolor[HTML]{FFF2CC}86.79\% & \cellcolor[HTML]{FFF2CC}38.00\% \\
\bottomrule
\end{tabular}%
}
\end{tiny}
\caption{Overall Pass@1 results. Light yellow cells mark the best score within each foundation model.}
\label{tab:results_pass1}
\end{table*}

For TALM’s tree-structured collaboration, we set the maximum depth to $m=3$, the initial branching factor to $n=3$, and the decay rate to $k=1$. The validation agent is allowed up to $r=3$ correction attempts. In the long-term memory module, the similarity threshold for new entries is set to $0.75$, and up to $3$ relevant past records are retrieved during reasoning. A memory merge strategy is applied to consolidate redundant entries when overlap is detected.  

For MapCoder, we follow the authors’ original hyperparameter settings: retrieval size and retry count are set to $5$ for HumanEval and $3$ for the other datasets.  

As the evaluation metric, we adopt Pass@1, which considers a task solved if any of the top-$1$ generated outputs is correct. For fairness, TALM’s memory is cleared at the beginning of each experimental round, and dataset orders are randomized to avoid bias from fixed learning trajectories.

\subsection{Main Results}
\label{sec:results}
\subsubsection{Performance on Code Generation}
Table~\ref{tab:results_pass1} reports Pass@1 performance across HumanEval, BigCodeBench, and ClassEval. Overall, TALM consistently outperforms baseline prompting methods and the multi-agent framework MapCoder. 

On the relatively simple HumanEval benchmark, all methods achieve strong accuracy, reflecting the limited challenge of single-function synthesis. Nevertheless, TALM delivers a modest but consistent improvement over direct prompting and CoT, demonstrating that even for short-context tasks, structured decomposition and validation provide benefits. 

For BigCodeBench, which requires handling long contexts and realistic engineering subtasks, most prompting-based methods exhibit limited gains and in some cases degrade performance. TALM, by contrast, achieves an improvement of roughly 2-3\% over the direct prompting baseline, highlighting its robustness under more demanding conditions. 

ClassEval presents the most challenging setting, requiring coherent class-level generation across multiple methods. At the testcase level, TALM improves by over 6\% compared to direct prompting, and by 2–3\% relative to the next-best baseline. At the class level, where structural consistency is critical, TALM achieves the highest score, outperforming other methods by around three percentage points. These results confirm that TALM’s tree-structured decomposition and localized correction mechanisms are particularly effective when tasks demand both modularity and long-horizon reasoning.

\subsubsection{Token Usage Efficiency}
\begin{figure}[t]
  \centering
  \includegraphics[width=0.7\linewidth]{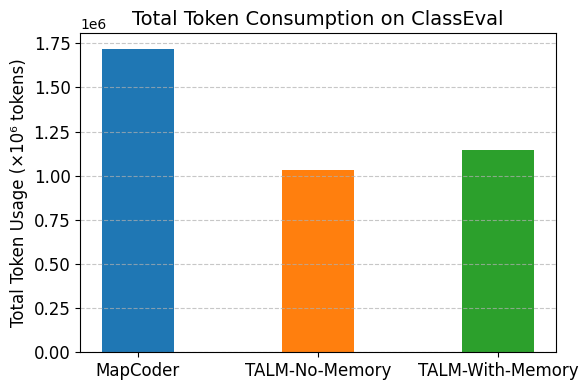}
  \caption{Comparison of token consumption between TALM and MapCoder.}
  \label{fig:results_tokens}
\end{figure}

Beyond accuracy, we also measure efficiency by tracking token consumption. Figure~\ref{fig:results_tokens} compares TALM with MapCoder, the other multi-agent framework baseline. TALM exhibits significantly lower token usage, both in input and output. This advantage stems from two design choices: (i) TALM’s simplified agent roles and hierarchical collaboration reduce the need to repeatedly pass large contexts across agents, and (ii) its long-term memory module retrieves past experiences directly from a vector store rather than generating long reference outputs during runtime. 

We further compare TALM with and without the long-term memory module. Results show only a marginal increase in token usage when memory is enabled, since retrieval and storage are handled externally by the vector database rather than the LLM itself. 
This indicates that TALM achieves accuracy gains with few additional token cost, making it well suited for large-scale, multi-round program synthesis tasks.

\subsection{Ablation Studies and Analyses}
\label{sec:ablation}
\subsubsection{Impact of Long-Term Memory}
We first examine the contribution of the long-term memory module from Table~\ref{tab:results_pass1}. While the overall Pass@1 improvement is modest on HumanEval, the gains are more pronounced on BigCodeBench and ClassEval, where tasks are longer and structurally more complex. This suggests that memory provides greater benefits in settings where knowledge reuse across tasks is meaningful. 

\begin{figure}[H]
  \centering
  \includegraphics[width=\linewidth]{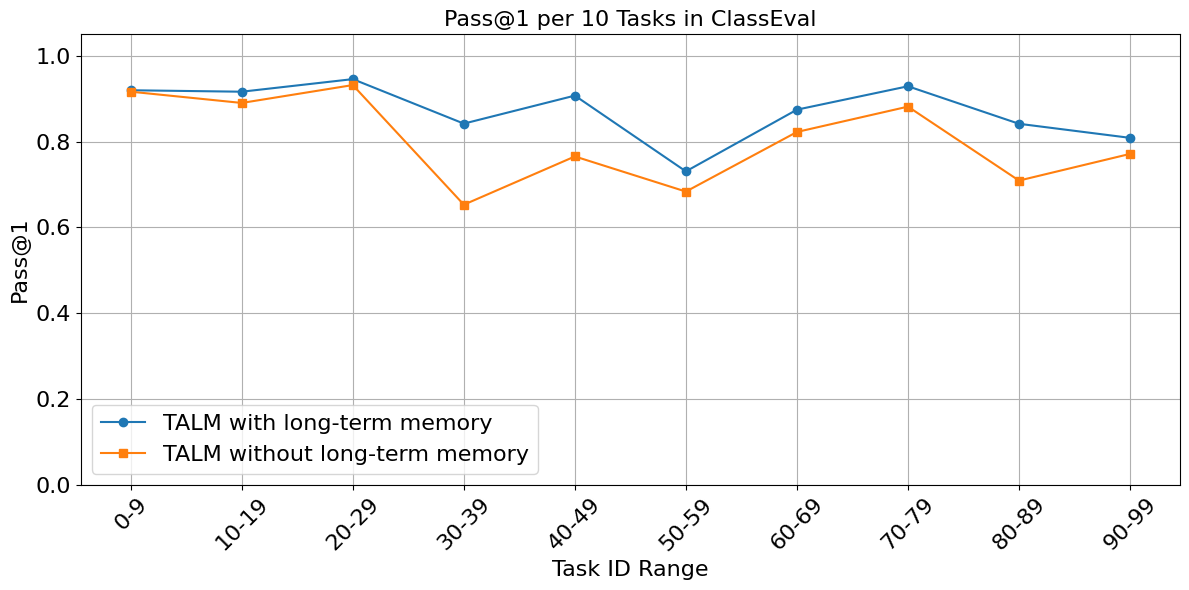}
  \caption{Effect of the long-term memory module over task sequence on ClassEval.}
  \label{fig:ablation_memory}
\end{figure}

Figure~\ref{fig:ablation_memory} further illustrates performance progression on ClassEval over the task sequence. During early problems, the memory store is still sparse, and performance remains close to baseline. As tasks progress, accumulated knowledge begins to provide tangible benefits, leading to a stable improvement in later segments. This result highlights the potential of the TALM framework for long-chain and structurally complex applications.

\subsubsection{Impact of Localized Re-Reasoning}
Table~\ref{tab:ablation_rerereasoning} reports results when disabling the re-reasoning mechanism. Across all datasets, removing localized re-reasoning leads to consistent performance drops, regardless of whether long-term memory is enabled. The effect is particularly notable on BigCodeBench and ClassEval, where error propagation is more common due to task complexity. These findings highlight the importance of localized re-reasoning: it allows TALM to correct reasoning flaws at the subtree level, avoiding costly restarts while preserving verified outputs.

\begin{table}[htbp]
\centering
\resizebox{\columnwidth}{!}{%
\begin{tabular}{lccc}
\hline
 & \textbf{HumanEval} & \textbf{BigCodeBench} & \textbf{ClassEval} \\ \hline
TALM (without memory) & \begin{tabular}[c]{@{}c@{}}89.34\%\\ (-1.25\%)\end{tabular} 
                      & \begin{tabular}[c]{@{}c@{}}45.11\%\\ (-1.21\%)\end{tabular} 
                      & \begin{tabular}[c]{@{}c@{}}75.19\%\\ (-2.66\%)\end{tabular} \\ \hline
TALM (with memory)    & \begin{tabular}[c]{@{}c@{}}86.42\%\\ (-1.51\%)\end{tabular} 
                      & \begin{tabular}[c]{@{}c@{}}44.92\%\\ (-2.18\%)\end{tabular} 
                      & \begin{tabular}[c]{@{}c@{}}78.44\%\\ (-2.28\%)\end{tabular} \\ \hline
\end{tabular}%
}
\caption{Ablation on localized re-reasoning. Numbers in parentheses indicate performance drop compared to full TALM.}
\label{tab:ablation_rerereasoning}
\end{table}

\subsubsection{Impact of Tree Height}
We next analyze the effect of varying the maximum tree height $m$ (Table~\ref{tab:ablation_treeheight}). Results indicate that performance improves as $m$ increases up to an optimal level, beyond which it declines. For HumanEval, the best result occurs at $m=2$, while for the more complex BigCodeBench and ClassEval datasets, $m=3$ is optimal. Further increases in $m$ lead to over-decomposition: agents tend to split tasks excessively, consuming resources and introducing unnecessary reasoning steps. This confirms that tree height is a critical hyperparameter balancing task granularity and reasoning efficiency.

\begin{table}[htbp]
\centering
\resizebox{\columnwidth}{!}{%
\begin{tabular}{cccc}
\hline
\rowcolor[HTML]{FFFFFF} 
\multicolumn{1}{l}{\cellcolor[HTML]{FFFFFF}Tree Height / Benchmark} & \textbf{HumanEval} & \textbf{BigCodeBench} & \textbf{ClassEval} \\ \hline
$m=1$ & 87.13\% & 45.55\% & 78.33\% \\
$m=2$ & \cellcolor[HTML]{FFF2CC}89.13\% & 46.19\% & 78.62\% \\
$m=3$ & 87.93\% & \cellcolor[HTML]{FFF2CC}47.10\% & \cellcolor[HTML]{FFF2CC}80.72\% \\
$m=4$ & 83.98\% & 45.65\% & 80.31\% \\
$m=5$ & 81.32\% & 42.69\% & 76.95\% \\
\hline
\end{tabular}
}
\caption{Effect of maximum tree height $m$ on TALM performance.}
\label{tab:ablation_treeheight}
\end{table}

\subsubsection{Impact of Branching Factor}

Finally, we vary the maximum branching factor $n$ (Table~\ref{tab:ablation_degree}). Unlike tree height, increasing $n$ does not significantly degrade performance, as agents dynamically decide whether to spawn child agents. However, when $n$ is too small, complex tasks may lack sufficient decomposition flexibility, leading to performance drops. On HumanEval, a small branching factor ($n=1$) is sufficient, whereas on BigCodeBench and ClassEval, larger values (around $n=3$) yield the best results. These findings suggest that $n$ should be tuned according to task complexity—minimal branching for simple problems and broader branching for complex reasoning.

\begin{table}[htbp]
\centering
\resizebox{\columnwidth}{!}{%
\begin{tabular}{cccc}
\hline
\rowcolor[HTML]{FFFFFF} 
\multicolumn{1}{l}{\cellcolor[HTML]{FFFFFF}Degree of Tree / Benchmark} & \textbf{HumanEval} & \textbf{BigCodeBench} & \textbf{ClassEval} \\ \hline
$n=1$ & \cellcolor[HTML]{FFF2CC}88.22\% & 46.66\% & 79.33\% \\
$n=2$ & 88.09\% & 46.11\% & 78.34\% \\
$n=3$ & 87.93\% & \cellcolor[HTML]{FFF2CC}47.10\% & \cellcolor[HTML]{FFF2CC}80.72\% \\
$n=4$ & 87.88\% & 47.05\% & 80.31\% \\
$n=5$ & 87.90\% & 47.08\% & 80.11\% \\
\hline
\end{tabular}
}
\caption{Effect of maximum branching factor $n$ on TALM performance.}
\label{tab:ablation_degree}
\end{table}

\section{Conclusions and Future Work}
This work introduced \textbf{TALM} (Tree-Structured Multi-Agent Framework with Long-Term Memory), a code generation framework designed to address the limitations of existing multi-agent collaboration systems in task flexibility, error correction, and knowledge retention. TALM advances the state of the art in three ways. First, it organizes agents into a dynamically expandable tree structure, allowing tasks to be decomposed and delegated without relying on rigid, pre-defined workflows. This enables the framework to adapt naturally to tasks of varying scale and complexity. Second, TALM introduces a localized re-reasoning mechanism, which confines error correction to the affected subtree rather than restarting the entire workflow. This significantly reduces computation overhead and enhances fault tolerance. Finally, TALM integrates a long-term memory module that stores prior problem–solution traces in a vector database, enabling retrieval of similar past experiences for future tasks and paving the way toward continual learning and implicit self-improvement.

We evaluated TALM on HumanEval, BigCodeBench, and ClassEval, comparing against both prompting-based baselines and the multi-agent framework MapCoder. Experimental results show that TALM consistently outperforms alternatives, particularly in complex and long-context settings. Ablation studies further confirmed the importance of localized re-reasoning and long-term memory, as well as the impact of tree parameters such as maximum depth and branching factor. Together, these findings highlight TALM’s robustness, efficiency, and alignment with real-world software engineering needs.

Our results also reveal a trade-off introduced by task decomposition. For relatively simple tasks, decomposition may introduce additional overhead or occasional failures due to increased token usage and compounding reasoning errors. In contrast, for more challenging problems, decomposition substantially improves solvability by breaking down complexity, effectively lowering the performance floor while raising the ceiling—an effect we consider more impactful in realistic software engineering scenarios.

Looking ahead, several promising directions emerge. First, TALM currently relies on fixed hyperparameters to constrain its tree structure. Future work may explore dynamic adjustment mechanisms, enabling agents to adaptively tune tree height and branching during runtime. Second, while the memory module is fully automated, its quality could be further enhanced by incorporating human feedback or lightweight verification before updating the knowledge base. Third, TALM has thus far been evaluated only on natural-language-to-code tasks. Extending it to other software engineering activities—such as code completion, refactoring, and debugging—could broaden its applicability and strengthen its utility as a general-purpose intelligent assistant.

\section{Limitations}
Despite its promising results, TALM has several limitations that should be acknowledged. First, the datasets used in this study are relatively modest in scale. HumanEval and ClassEval contain fewer than two hundred problems each, and even BigCodeBench consists of only 1,140 problems. This restricts TALM’s ability to fully demonstrate the long-term benefits of its memory module. Moreover, to ensure fairness, the memory was reset at the start of each experimental round, which further underrepresents its potential in real-world deployments where memory would accumulate continuously.

Second, while ClassEval poses greater contextual and structural challenges than HumanEval or BigCodeBench, it remains limited to class-level tasks. In practice, software engineering often involves system-level projects spanning multiple classes, modules, or packages. Such large-scale contexts were not available in existing benchmarks, leaving open the question of how TALM would perform on truly large and interconnected codebases.

In summary, TALM demonstrates notable improvements in flexibility, error recovery, and knowledge reuse compared to prior MAC frameworks. However, its evaluation has been constrained by dataset scale and scope. Overcoming these limitations will be critical to realizing TALM’s full potential and advancing toward autonomous, self-improving agents for software development.

\bibliographystyle{ACM-Reference-Format}
\bibliography{references}

\appendix
\section{Code Generation Sample by TALM}
\label{app:code-sample}
Here we provide a representative code sample generated by TALM while handling a task from BigCodeBench. The example demonstrates TALM’s ability to decompose problems and generate structured, executable code, offering qualitative insight beyond the quantitative evaluation reported in the main paper.

\begin{lstlisting}
\begin{small}{
import cgi
import http.server
import smtplib
from email.mime.text import MIMEText
import json

def task_func(smtp_server, smtp_port, smtp_username, smtp_password):
    REQUIRED_KEYS = ("subject", "message", "to")

    class EmailRequestHandler(http.server.BaseHTTPRequestHandler):
        server_version = "EmailHTTP/1.0"
        sys_version = ""

        def do_POST(self):
            try:
                content_type = self._require_json_content_type()
                body_bytes = self._read_request_body()

                email_data = self._parse_json(body_bytes)
                self._validate_email_data(email_data) 

                msg = self._build_mime_message(email_data)
                self._send_email(msg, email_data["to"])

                self._send_ok()
            except json.JSONDecodeError:
                self._send_error_response(400, "Bad Request: invalid JSON")
            except ValueError as e:
                self._send_error_response(400, f"Bad Request: {e}")
            except smtplib.SMTPAuthenticationError:
                self._send_error_response(535, "Authentication Failed")
            except Exception:
                self._send_error_response(500, "Internal Server Error")

        def _require_json_content_type(self) -> str:
            raw = self.headers.get("content-type")
            if not raw:
                raise ValueError("Missing Content-Type header")

            ctype, _pdict = cgi.parse_header(raw)
            if ctype.lower() != "application/json":
                raise ValueError("Content-Type must be application/json")
            return ctype

        def _read_request_body(self) -> bytes:
            raw_len = self.headers.get("content-length")
            if raw_len is None:
                raise ValueError("Missing Content-Length header")

            try:
                length = int(raw_len)
            except ValueError:
                raise ValueError("Invalid Content-Length header")

            if length < 0:
                raise ValueError("Invalid Content-Length header")

            return self.rfile.read(length)

        def _parse_json(self, body: bytes) -> dict:
            text = body.decode("utf-8")
            data = json.loads(text)
            if not isinstance(data, dict):
                raise ValueError("JSON body must be an object")
            return data

        def _validate_email_data(self, data: dict) -> None:
            missing = [k for k in REQUIRED_KEYS if k not in data]
            if missing:
                raise ValueError(f"Missing required key(s): {', '.join(missing)}")

            if not isinstance(data["subject"], str) or not data["subject"].strip():
                raise ValueError("'subject' must be a non-empty string")
            if not isinstance(data["message"], str):
                raise ValueError("'message' must be a string")
            if not isinstance(data["to"], str) or "@" not in data["to"]:
                # Not full RFC validation; avoids obvious garbage
                raise ValueError("'to' must be a valid email-like string")

        def _build_mime_message(self, data: dict) -> MIMEText:
            msg = MIMEText(data["message"], _subtype="plain", _charset="utf-8")
            msg["Subject"] = data["subject"]
            msg["From"] = smtp_username
            msg["To"] = data["to"]
            return msg

        def _send_email(self, msg: MIMEText, recipient: str) -> None:
            with smtplib.SMTP(smtp_server, smtp_port, timeout=20) as server:
                server.ehlo()
                try:
                    server.starttls()
                    server.ehlo()
                except smtplib.SMTPException:
                    pass

                server.login(smtp_username, smtp_password)  # may raise SMTPAuthenticationError
                server.sendmail(smtp_username, [recipient], msg.as_string())

        def _send_ok(self) -> None:
            body = b"OK"
            self.send_response(200)
            self.send_header("Content-Type", "text/plain; charset=utf-8")
            self.send_header("Content-Length", str(len(body)))
            self.end_headers()
            self.wfile.write(body)

        def _send_error_response(self, code: int, message: str) -> None:
            body = message.encode("utf-8")
            self.send_response(code)
            self.send_header("Content-Type", "text/plain; charset=utf-8")
            self.send_header("Content-Length", str(len(body)))
            self.end_headers()
            self.wfile.write(body)

        def log_message(self, fmt, *args):
            return

    return EmailRequestHandler
}\end{small}
\end{lstlisting}

\end{document}